\documentclass[aps,prb,twocolumn,superscriptaddress]{revtex4}
\usepackage{graphicx}
\usepackage{epsfig}
\usepackage{bm}
\usepackage{times}
\usepackage{amssymb}
\usepackage{url,hyperref}

\begin{document}

\title{Variational study of the antiferromagnetic insulating phase of V$_2$O$_3$
 based on Nth order Muffin-Tin-Orbitals.}

\author{N. B. Perkins}
\affiliation{University of Wisconsin - Madison, 1150 University Avenue Madison, WI 53706-1390, USA}
\author{S. Di Matteo}
\affiliation{Equipe de Physique des Surfaces et Interfaces, Institut de Physique de Rennes UMR CNRS-UR1 6251, Universit\'e de Rennes1, F-35042 Rennes (France)}
\author{C.R. Natoli}
\affiliation{Laboratori Nazionali di Frascati, INFN, Casella Postale 13, I-00044 Frascati, Italy}

\date{\today}

\begin{abstract}
 Motivated by recent results of $N$th order muffin-tin
orbital (NMTO) implementation of the density functional theory
(DFT), we re-examine low-temperature ground-state properties of the
anti-ferromagnetic insulating phase of vanadium sesquioxide
V$_2$O$_3$. The  hopping matrix elements obtained by the
NMTO-downfolding procedure differ significantly from those previously
obtained in electronic structure calculations and imply that the in-plane
hopping integrals are as important as the out-of-plane ones. We use the NMTO hopping
matrix elements as input and perform a variational study of  the ground state.
We show that the formation of stable  molecules throughout the crystal is not
favorable in this case and that the experimentally observed
magnetic structure can still be obtained in the atomic variational regime.
However the resulting ground state (two $t_{2g}$ electrons occupying the
degenerate $e_g$ doublet) is in contrast with many well established
experimental observations. We discuss the implications of this finding
in the light of the non-local electronic correlations certainly present in
this compound.

\end{abstract}
\maketitle

\section{Introduction}

Vanadium sesquioxide  V$_2$O$_3$  holds a  a very
special place among the
 variety of physical systems exhibiting a metal-insulator
transition (MIT). This compound is considered as a prototype of
a Mott-Hubbard system: it displays a MIT
from a paramagnetic metallic (PM) phase to an antiferromagnetic
insulating (AFI) phase at low temperatures ($\approx 150 K$), and
a transition from a PM phase to a paramagnetic insulating (PI) phase at a
higher temperature ($\approx 500 K$).\cite{rice69,mcwhan70,dernier70,moon70}
In order to understand several of its properties, the knowledge of its crystal structure plays a fundamental role : in the paramagnetic phase
V$_2$O$_3$ can be characterized by a corundum cell in which  V ions
are arranged in V-V pairs along the $c$-hexagonal axis and form a
honeycomb lattice in the basal $ab$ plane. All vanadium ions are equivalent: each V$^{3+}-$ion has $3d^2$
configuration and is surrounded by a nearly perfect oxygen octahedron. However, a slight displacement
of vanadium ions away from the centers of their octahedra is at the origin of a small trigonal distortion which removes the
three-fold degeneracy of the $t_{2g}$ manifold. A non-degenerate $a_{1g}$  orbital and doubly degenerate $e_g$
orbitals are then separated by an energy gap measuring the trigonal distortion, $\Delta_t$. A first
order  structural transition takes place upon cooling, and the
system becomes monoclinic. Simultaneously,  a peculiar
antiferromagnetic (AF) spin order emerges with  ferromagnetically
ordered planes antiferromagnetically stacked perpendicularly to the monoclinic $b_m$ axis.

In late seventies, Castellani, Natoli and Ranninger \cite{cnr} (CNR)
developed the first realistic description of V$_2$O$_3$. They
realized that the peculiar structure observed in the AFI phase could
not be explained in terms of a single band Hubbard model, and that
the introduction of the orbital degrees of freedom into the
model was a necessary ingredient in order to explain the experimental findings. The CNR
model has been considered as a reliable model of V$_2$O$_3$ until the end of the nineties, when several experiments, e.g.,
x-ray absorption spectroscopy by Park et al.\cite{park00} and non-resonant magnetic x-ray scattering by Paolasini et al. \cite{paolasini}
have demonstrated that this model was in need of corrections. The failing of CNR
model came from the underestimation of the value of Hund's coupling ($J\sim 0.2\div 0.3$ eV) and the subsequent result
of a spin-state $S=1/2$ for $V^{3+}-$ions.
 Instead, all previous experiments independently showed that each vanadium ion has a spin $S=1$, observation that called for a deep revision of the theoretical description
of  V$_2$O$_3$. For this reason, many theoretical works followed.\cite{ezhov,mila,prb2002,elfimov,tanaka,laad,poteryaev}
In particular, very recently, the nature of the MIT in V$_2$O$_3$ has been studied by means of
LDA+DMFT approaches.  Laad {\it et al}\cite{laad} have
 proposed that the MIT is of
orbital-selective type, and that large changes in the orbital
occupation at the transition  occur in response to the
 crystal field changes. Poteryaev {\it et
 al}\cite{poteryaev}, with a different LDA+DMFT approach showed  that the crystal field, the correlations,
 and the orbital degrees
of freedom are strongly interrelated. They
also proposed that the trigonal distortion, strongly enhanced by
correlations, acts as an external field in the orbital
Hilbert space, in the same way as a magnetic field acts in the spin space.
These two works\cite{laad,poteryaev} have suggested that
 drastic modifications in the
hybridization of the orbitals due to the correlation effects and the
trigonal splitting were at the basis of the change in the population of
the orbitals across the transition.

Despite all this progress in the understanding of  the MIT in
V$_2$O$_3$, the controversy about the nature of the low-temperature
antiferromagnetic phase remains. Specifically, there are contradictory results on  how the two electrons  are distributed within the
$t_{2g}$ manifold, and whether or not orbital ordering takes place.
The LDA+U calculation of Ezhov {\it et al.} \cite{ezhov} suggested
that the $S=1$ state with no orbital degeneracy and a purely $e_g$ occupancy is a possible candidate
 for the low-temperature antiferromagnetic phase. This idea, {\it mutatis mutandis}, is present also in the LDA+DMFT calculations of Ref. [\onlinecite{poteryaev}].
Mila {\it et al.} \cite{mila} on the contrary used an old idea by Allen \cite{allen76}
that  magnetic and optical properties of all the phases of
V$_2$O$_3$ show a loss of $V^{3+}$-ion identity, and suggested that
a good candidate for a ground state is a state in which vanadium
ions have $S=1$ and form vertical bonds (molecules) with the total
spin $S_{tot}=2$. In our previous  study\cite{prb2002},  we showed using a different
correlated model that the molecular state can be stabilized
throughout the crystal only if the correlation energy within the molecule is big compared to the in-plane interaction energy. This gave
a certain region  in the parameter space where the molecular
solution could be realized. Later, Tanaka \cite{tanaka} proposed that,
in order to explain a large contribution from the orbital magnetic
moment, $<L> ~\sim -0.5 \mu_B$ (the minus sign reminds that this contribution is opposite to the spin contribution $2<S> ~\sim 1.7 \mu_B$), observed in
the non-resonant magnetic scattering experiments by Paolasini {\it
et al.} \cite{paolasini}, the orbital wave function had to be complex.
He argued that spin-orbit coupling plays an important role in
determining both the ground state and the low energy  excitation
spectrum of V$_2$O$_3$.

This controversy of theoretical results calls
for a better quantitative description of the electronic structure of
V$_2$O$_3$, which  is a starting point for any model-based analysis.
Recently,  the NMTO  method has been applied to study the electronic
structure of V$_2$O$_3$.\cite{andersen} Surprisingly, after early
works by CNR\cite{cnr} and by Mattheis,\cite{mattheiss} this is the first
serious attempt to model  complicated electronic structure of
V$_2$O$_3$ by an effective single particle Hamiltonian. The NMTO
method provides a way to derive an effective Hamiltonian with only
few energy-selective Wannier-like orbitals by integrating out all
others, less important degrees of freedom.  The result is
 an accurate tight-binding (TB) modeling of V$_2$O$_3$. The NMTO
approach   shows that oxygen contribution to hopping matrix
elements and renormalization effects due to hopping paths via the
$e_g$-tails (in addition to direct V-V $t_{2g}$-hopping paths)
 are  important and  were only partially taken into
account in Refs. [\onlinecite{mattheiss,cnr}]. As one can see from
 Table I, the TB hopping parameters obtained
by the NMTO-downfolding technique differ significantly from
 the estimates based on the TB fitting performed on the LAPW band
calculation of Mattheiss\cite{mattheiss} or by CNR.
 The NMTO analysis shows that the hopping through $e_g$-tails mainly
influences the parameters in the basal plane. This is a
 bonding effect, which leads to an increase of the in-plane matrix elements.
 The other
contributions via oxygen tails affect mainly the hopping along
vertical bonds. This is an anti-bonding effect which significantly
reduces $a_{1g}-a_{1g}$ hopping matrix elements.
 The latter leads to the decrease of the correlation
energy of vertical pairs, in this way reducing the stability of the
molecular state. Simultaneous increase of the in-plane interaction
energy favors uncorrelated atomic sites, and  in the end potentially breaks the
molecular state. This new qualitative picture of the ground
state of V$_2$O$_3$ in the parameter space, suggested by
NMTO-downfolding method,\cite{andersen} calls for re-examination of
the  variational analysis of the effective Hamiltonian in the
antiferromagnetic insulating state, which is the main purpose of
this paper.

The structure of the  paper is  as follows. We first present a brief outline
of the results of our previous work, Ref. [\onlinecite{prb2002}] and  set the notations.
We then minimize the effective Hamiltonian we derived using the values of the hopping parameters from
Ref. [\onlinecite{andersen}] and performing a variational
calculation with two kinds of variational functions: molecular and atomic.
We discuss the result in the light of the experimental facts and try to draw
some conclusions for the physics of V$_2$O$_3$.

\vspace{5mm} \noindent {\bf Table I}. Transfer integrals (eV) from TB
calculations used by CNR\cite{cnr}, from LAPW-calculations by
Mattheiss\cite{mattheiss}, and from NMTO-downfolding technique by
Saha-Dasgupta\cite{andersen}. Notations are explained in Table II in
the Ref. [\onlinecite{prb2002}].
\begin{center}
\begin{tabular}{|c|c|c|c|}\hline
 & Castellani {\em et al.}\cite{cnr} & Mattheiss\cite{mattheiss} & Saha-Dasgupta\cite{andersen}
\\\tableline
$\mu$ & $0.2$ & $0.2$ &   0.06 \\\tableline
$\rho$ & $-0.72$ & $-0.82$& -0.51\\\tableline
$-\alpha$ & $-0.13$ & $-0.14$& 0.08 \\\tableline
$\beta$ & $-0.04$ & $-0.05$& -0.21  \\\tableline
$\sigma$ & $0.05$ & $0.05$ &  -0.03\\\tableline
$-\tau$ & $-0.23$ & $-0.27$&  -0.26\\\hline
\end{tabular}
\end{center}

\section{Model, formalism and notations}

We assume that the low-temperature AFI phase of V$_2$O$_3$ can be described by a super-exchange
spin-orbital Hamiltonian, $H_{\rm SE}$, that can be derived perturbatively from a three-fold degenerate $t_{2g}$
Hubbard Hamiltonian, $H_H$, as done in Ref. [\onlinecite{prb2002}], to which we refer for further details. We can write the Hubbard Hamiltonian as:
\begin{eqnarray}
 H_H=H_{\rm t}+H_{\rm U}~,
\end{eqnarray}
\noindent where the kinetic term $H_{\rm t}=\sum_{jj'}\sum_{mm'
\sigma }t_{jj'}^{mm'} c^{+}_{jm \sigma }c_{j'm' \sigma }$  includes
a summation over nearest neighbor (nn) sites, over  orbital
($m,m'=1,2,3$) and spin ($\sigma=\uparrow,\downarrow$) indexes.
$H_{\rm U}$ describes  the on-site Coulomb interactions $U_1$ (for electrons in the same orbital) and $U_2$ (for electrons in different orbitals) and
Hund's coupling $J$. The hopping integrals $t_{jj'}^{mm'}$ ($m,m'=1,2,3$) can be
expressed via a reduced set of parameters: $\mu$ and $\rho$ for out-of-plane hopping (ie, within the molecule), $\alpha$,
$\beta$, $\sigma$ and $\tau$ for in-plane hopping. In  Table I we present the numerical values of hopping matrix
elements obtained by CNR\cite{cnr} through unrestricted Hartree-Fock calculations, by Mattheiss\cite{mattheiss} throug a fitting procedure of the LAPW-calculations, and by
Saha-Dasgupta et al\cite{andersen} through the NMTO-downfolding technique. We shall later use these parameters
in the variational analysis where we compare various ground-states energies.
For the Coulomb repulsion $U_2$ and Hund's coupling $J$
Ezhov {\it et al.}\cite{ezhov} and Mila {\it et al.}\cite{mila} suggested
$J\sim 1.0$ eV , $U_2\sim 2.5$ eV. Recent optical studies of
Qazilbash {\it et al}\cite{basov}, however, yielded a smaller
value for the Hund's coupling $J=0.5$ eV.  We have described already\cite{prb2002} how the range of these two parameters is subjected to large fluctuations in the literature. In what follows we shall fix $U_2 \sim 2.5$ eV and
 consider a range of $J\simeq 0.4 \div~ 1.0$ eV .

We further assume, as found experimentally, that at each site two $t_{2g}$-electrons are bound
into a $S=1$ state. We therefore derive from $H_{\rm H}$ an effective
Hamiltonian, $H_{\rm eff}=H_{\rm SE}+H_{\rm trig}$ to describe the insulating phase of V$_2$O$_3$, where $H_{\rm SE}$ and  $H_{\rm trig}$
are given by
\begin{eqnarray}
\begin{array}{lll}
H_{\rm SE} & = & -\frac{1}{3}\frac{1}{U_2-J}\sum_{ij}{\big [}
2+\vec S_i\cdot \vec S_j {\big ]}O^{(1)}_{ij}\\[0.3cm]
 & & -\frac{1}{4}\frac{1}{U_2+4J}\sum_{ij}{\big [}
1-\vec S_i\cdot \vec S_j {\big ]}O^{(2)}_{ij}\\[0.3cm]
 & & -\frac{1}{12}\frac{1}{U_2+2J}\sum_{ij}{\big [}
1-\vec S_i\cdot \vec S_j {\big ]}O^{(3)}_{ij}\\
\end{array}
 \label{simpform}
\end{eqnarray}
\begin{eqnarray}
\begin{array}{lll}
H_{\rm trig} & = & +\sum_{j m \sigma }
\Delta_{m} n_{jm\sigma},
\end{array}
 \label{simpform2}
\end{eqnarray}
Here $S_j=1$ is the spin at site $j$ and $n_{jm\sigma}$ describes the occupation of the $m$ orbital on site $j$
by an electron with spin $\sigma$, whereas the $O^{(k)}_{ij}$ are  orbital
exchange operators presented in Appendix C of Ref. [\onlinecite{prb2002}].
Multiplied by the corresponding  prefactors $-\frac{1}{3}\frac{1}{U_2-J}$,
$-\frac{1}{4}\frac{1}{U_2+4J}$ or $-\frac{1}{12}\frac{1}{U_2+2J}$,
they define the effective value of the exchange that depends on the orbital
occupation of the two sites $i$ and $j$ along the bond direction $<ij>$.
The spin terms can be simply considered as projectors on the FM and AFM
state. The term $H_{\rm trig}$ describes the trigonal distortion that
splits the three  degenerate two-electron states: $|e_g^1 e_g^2\rangle
\: \equiv |0\rangle$, $|a_{1g}e_g^1\rangle \: \equiv |-1\rangle$,
$|a_{1g}e_g^2\rangle \: \equiv |1\rangle$.
The energy splittings due to this distortion are defined as
$\Delta_{1} = \Delta_{2} = 0$ and $\Delta_{3} = \Delta_t
> 0$  for orbitals $m=1,2$ and 3, respectively.
$\Delta_t$ is comparable in magnitude to the hopping integrals,
 which allows us to treat the term $H_{\rm trig}$ on the same level as $H_{\rm SE}$.

\section{Results}

In this section, we compare  orbital
and magnetic ground-state configurations of the effective
Hamiltonian $H_{\rm eff}$ as obtained  by a variational analysis
with TB parameters from
Mattheiss\cite{mattheiss}  and NMTO\cite{andersen} models.

The trial wave function can be most generally written as
\begin{equation}
|\Psi\rangle =\Pi_{n}~|\Psi_n\rangle =\Pi_{n}~|\psi^o_n\rangle
|\phi^s_n\rangle \label{variational}
\end{equation}
where $|\psi^o_n\rangle$ describes the orbital part and
$|\phi^s_n\rangle$ refers to the spin part of the wave function  on
a site $n$.

 In the
following, we shall use as a variational wave function
$|\Psi_n\rangle$ either an atomic state or a molecular state, with $n$
labeling an atomic or a molecular site, respectively.
 In both cases,
the average value  of the $H_{\rm eff}$ over the corresponding state
takes the form:
\begin{eqnarray}
\begin{array}{c}
\langle\Psi_n|\langle\Psi_m|H_{\rm eff}|\Psi_m\rangle |\Psi_n\rangle
=
\\[0.2cm]
\langle\psi^o_n|\langle\psi^o_m|H^o_{\rm eff}|\psi^o_m\rangle
|\psi^o_n\rangle \times \langle\phi^s_n|\langle\phi^s_m|H^s_{\rm
eff}|\phi^s_m\rangle |\phi^s_n\rangle
\end{array}
\label{av}
\end{eqnarray}
This factorization is possible only in the mean field approximation,
in which single-site wave-functions, $|\Psi_n\rangle$, otherwise entangled,
factor into orbital $|\psi^o_n\rangle$ and spin $|\phi^s_n\rangle$ parts.
By this we explicitly neglect any coupled spin-orbit
fluctuations. The spin averaging  is straightforward: for a
ferromagnetic bond $\langle \vec{S}_n\cdot\vec{S}_m+2\rangle_{HF}=3$
and $\langle \vec{S}_n\cdot\vec{S}_m-1\rangle_{HF}=0$, while for an
antiferromagnetic bond, $\langle
\vec{S}_n\cdot\vec{S}_m+2\rangle_{HF}=1$ and $\langle
\vec{S}_n\cdot\vec{S}_m-1\rangle_{HF}=-2$. Orbital averaging in
Eq.(\ref{av}), however, requires some algebra
 and is discussed in  Ref. [\onlinecite{prb2002}] for both atomic
and molecular states. In the following, we will first average over
all possible ordered magnetic structures, and then for each type of
magnetic ordering find an orbital configuration that minimizes the
total energy be means of a variational procedure. Note, that we
limit the number of magnetically ordered structures only to those
that can be realized on the corundum unit cell.

As a preliminary step, before the full minimization of the variational energy (\ref{av}), we can compare the correlation energy
of the ferromagnetic state of the vertical pair, $2\rho\mu/(U_2-J)$,
 with the superexchange energy in the basal plane, in order to have an idea about the relative order of magnitude of atomic and molecular energies. We remind that the correlation energy is
defined as the difference between the exact ground state energy,
$E_V=-(\rho-\mu)^2/(U_2-J)$, and the ground state energy in the
Hartree-Fock approximation, $E_V^{HF}-(\rho^2+\mu^2)/(U_2-J)$. The superexchange energy in the basal plane was approximated
by $(\alpha^2+\tau^2)/(U_2-J)$ in  Ref. [\onlinecite{prb2002}],
 where we  assumed that the contributions from $\beta$ and $\sigma$
 hopping matrix elements were negligible, and should rather be written as $(\beta^2+\tau^2)/(U_2-J)$ with the NMTO values of Table I, that shows that $\beta$ is no more negligible, whereas $\alpha$ it is.

There are two qualitatively different regimes of solutions. If the
correlation energy is larger,  the most appropriate variational
wave function for the whole $H_{\rm eff}$ must be constructed in
terms of molecular units with  $S^M_z=2$  and orbital part
$|\psi^o_n\rangle$ given by
\begin{eqnarray}\label{molecularwf}
|\psi^o_{\pm} \rangle_{ab}=\frac{1}{\sqrt{2}}(|\pm1 \rangle_a|0
\rangle_b+|\pm1 \rangle_b|0 \rangle_a) ~,
\end{eqnarray}
where $a$ and  $b$ define two sites of the vertical molecule, and $\pm 1$ and 0 denote  the value of $z$
component of the pseudospin operator, defining the orbital state.

If, instead,  the values of the exchange energy in the basal plane are
larger than the correlation energy, the appropriate variational
wave function is atomic-like and should be written as
\begin{eqnarray}\label{atomicwf}
|\psi^o_i\rangle =\cos\theta_i|0\rangle_{i}+\sin\theta_i
(\cos\psi_i|1\rangle_{i}+\sin\psi_i|-1\rangle_{i}) ~.
\end{eqnarray}
\noindent
Contrary to the molecular case, this wave function allows all
three states $|0\rangle_{i}$,
$|1\rangle_{i}$ and $|-1\rangle_{i}$ to be present without any {\it
a priori} restriction on their relative weight.
The relative weight of these three states is then
determined through the minimization procedure with respect to the
variational parameters $\theta_i$ and $\psi_i$.

As quantitatively shown in  Ref. [\onlinecite{prb2002}], Eqs. (6.19) and (6.20), in order to determine the nature (atomiclike, molecularlike) of the ground state, we need to compare the absolute value of the ratio of the molecular correlation energy (per atom) and the
in-plane exchange energy, computed  with Mattheiss parameters and
 new NMTO set of parameters.
We  obtain $\rho\mu/(\alpha^2+\tau^2)=1.7$ and $0.26$ for Mattheiss  and NMTO
set, respectively.  The strong reduction of this ratio for NMTO set
suggests that NMTO parameters do not favor the
formation of the molecular state.

There are few possible reasons
 for such large difference in the results of the two approaches.
 The discrepancy may be the consequence of the fact that
  NMTO method more accurately treats the effect of  trigonal distortion on
 hopping matrix elements. In the approach of Mattheiss
the influence of the trigonal distortion on matrix elements was assumed
 to be small and was neglected.
As one can see from Fig.14 of Ref. [\onlinecite{andersen}], where structures
with varying amount of distortion have been generated, the
$a_{1g}-a_{1g}$ hopping along vertical bond, $\rho$, decrease
significantly with the increase of the trigonal distortion: NMTO estimates
are $\rho (\Delta_t=0)=-1.2$ eV, while $\rho (\Delta_t=0.27{\rm
eV})=-0.51$ eV. The same figure gives much smaller estimate of
variation of $\rho$ for CNR model: $\rho (\Delta_t=0)=-0.8$ eV and
$\rho (\Delta_t=0.27{\rm eV})=-0.7$ eV. Other reasons of
discrepancy, as pointed by the authors of Ref. [\onlinecite{andersen}]
themselves, can be traced down to various approximations  used in
constructing  effective orbitals in the NMTO-downfolding,
and to imprecise knowledge of the value of the covalency
V($d$)-O($p$) mixing parameter and the charge transfer  $3d-2p$ gap,
extracted from nuclear magnetic resonance and photo-emission
experiments.


We now perform the minimization of the  effective
Hamiltonian with NMTO set of parameters and compare the results with the equivalent minimization procedure performed in Ref. [\onlinecite{prb2002}] with Mattheiss' parameters.
As in Ref. [\onlinecite{prb2002}], we  examine the following four magnetic phases:
\begin{itemize}
\item
AFM  phase --  all three in-plane bonds are antiferromagnetic;
\item
RS phase --
one in-plane bond is ferromagnetic and the other two are
antiferromagnetic;
 (this is the spin structure experimentally observed in V$_2$O$_3$);
\item
ARS phase --
one in-plane bond is antiferromagnetic and the other two are ferromagnetic;
\item
FM phase --  all three in-plane bonds are ferromagnetic.
\end{itemize}

\begin{figure}
     \epsfysize=60mm
     \centerline{\epsffile{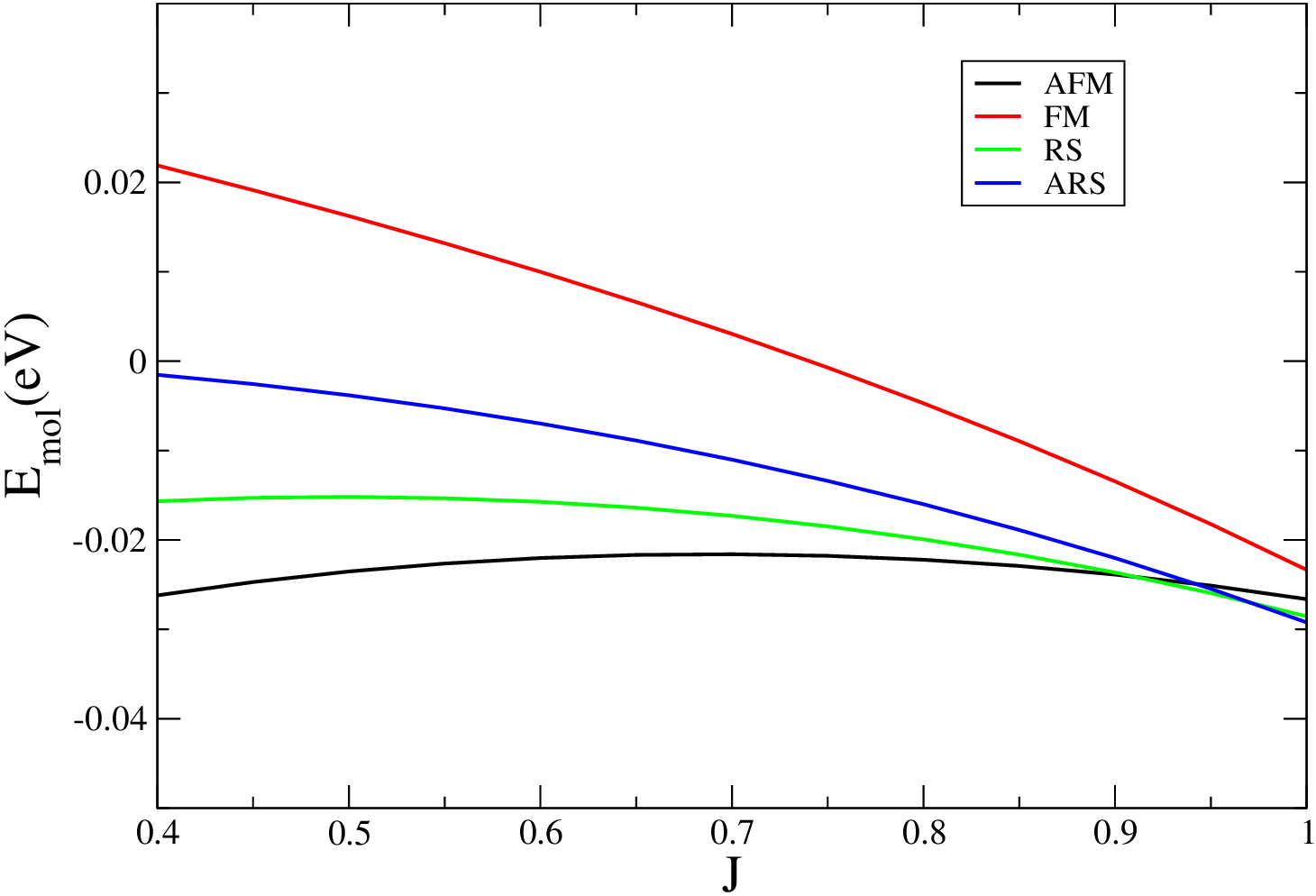}}
\vspace{1.5cm}
 \epsfysize=60mm
 \centerline{\epsffile{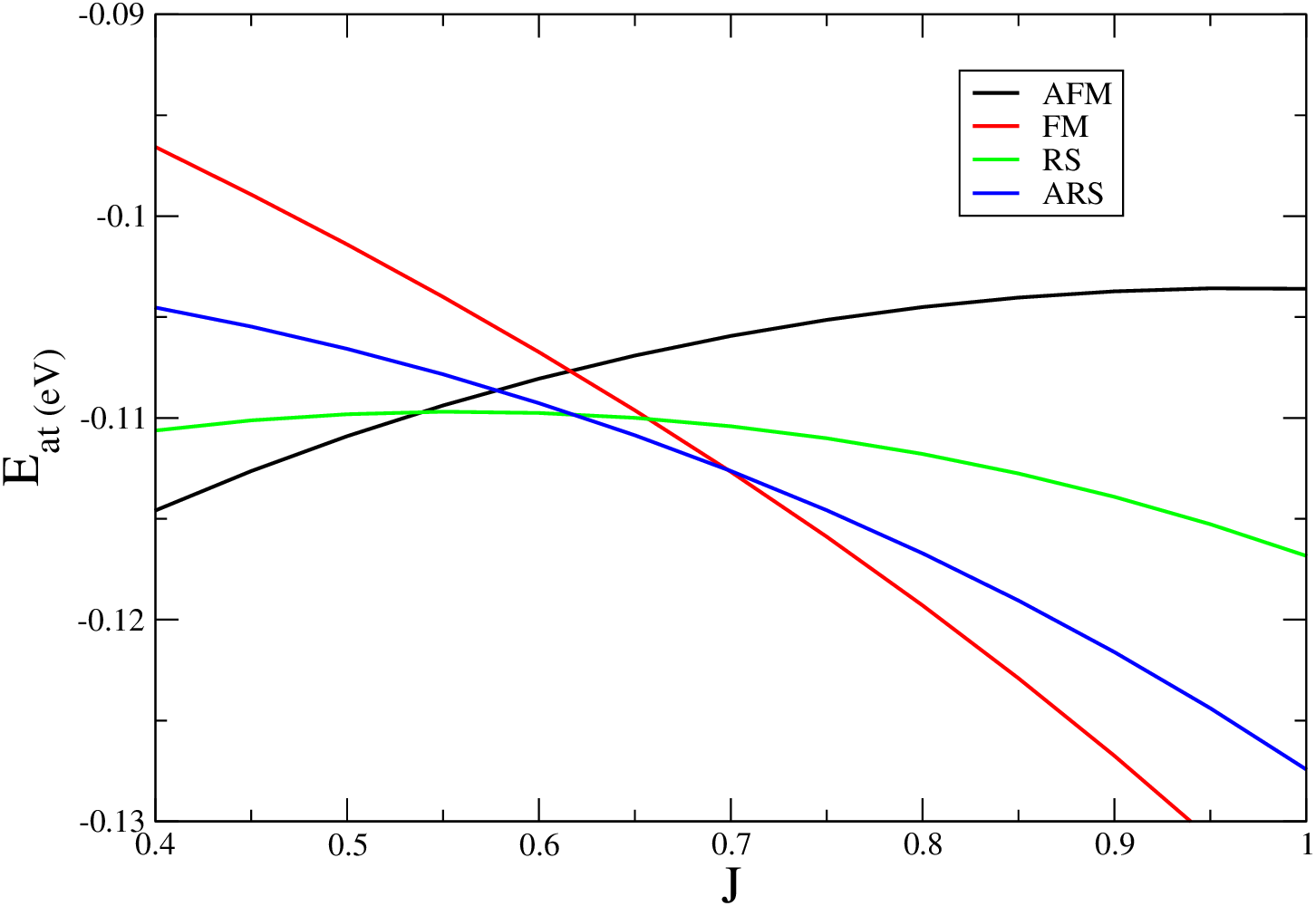}}
\caption{The energy per V atom as a function of $J$ for different
spin configurations (AFM, RS, ARS and FM type).  The hopping
parameters are defined by NMTO method\cite{andersen}, trigonal
splitting $\Delta_{t}=0.27$eV. In panel (a) the minimization is
performed with molecular trial wave function (\ref{molecularwf}),
while in panel (b) the minimization is performed with atomic trial
wave function (\ref{atomicwf}). (Color online)}\label{and}
\end{figure}

Fig.\ref{and} shows the plot of energy per V-ion, $E_V$, obtained
with molecular (a) and atomic (b) variational functions using the
NMTO set of parameters,
 as a function of Hund's coupling $J$ and for $U_2=2.5$eV.
A direct comparison between Fig. \ref{and} (a)
 and Fig. \ref{and} (b) shows that for all values of $J$
the energy of the atomic state is lower than the  energy of the
molecular state. This confirms what we had more qualitatively suggested
above -- that the
 molecular state is not supported by the
NMTO calculations.

The region of stability of experimentally observed RS magnetic
structure is again rather small, as  we have obtained previously
with Mattheiss parameters for the molecular state.
In this case, this region also shifts
 towards a lower value of Hund's coupling, $J\simeq 0.54
\div~ 0.62$eV. This value is  close to $J=0.5$ eV,
extracted from optical studies of Qazilbash {\it et al.}\cite{basov}, as well as to the old value $J=0.59$ eV given by Tanabe and Sugano\cite{sugano} by fitting optical spectra and at odds with the much bigger values ($J \sim 1$ eV) recently used in the literature.\cite{ezhov}

For the orbital structure, the  minimization gives the
 values of the orbital mixing angles $\theta_i\simeq 0$
and $\psi_i\simeq \pi/2$ at all sites for the RS configurations,
 ie, an orbital wave function $|\psi^o_i\rangle
=|0\rangle_{i}\equiv|e_g^1 e_g^2\rangle$.
 This is the solution without orbital degeneracy as found by
Ezhov {\it et al.}.\cite{ezhov} The absence of $a_{1g}$ electrons in
the solution can be understood as follows:  the small gain in the
kinetic energy, due to the significant reduction of
 the hopping matrix elements in the NMTO calculations, can not
compensate the effect of the trigonal splitting, which pushes
$a_{1g}$ orbital to higher energies and favors the occupancy of the
$|0\rangle$ states on all the atoms. As a result, the ground state
is a non-degenerate $e_g$-doublet, and the orbital degeneracy is
completely lifted, contrary to the ground states found in
Ref. [\onlinecite{mila,prb2002}], where the important $a_{1g}$ contribution to the kinetic
energy in the molecule due to the $a_{1g}-a_{1g}$ hopping term $\rho$ allowed a finite occupancy of $a_{1g}$ orbitals. It is
worth mentioning that even with Mattheiss set of parameters it was possible to find a ground state $|e_g e_g\rangle$,
however only for larger values of the trigonal splitting $\Delta_t>0.4$ eV.

One  more remark: due to the competing presence of the AFM, ARS and
FM phases, the stabilization energy  of the  RS atomic
solution is rather small, the first excited  state is located about 2 meV
above the ground state. Despite all differences between the ground
state solutions  obtained with NMTO and Mattheiss parameters, this
is a common  feature which is apparent from Fig.\ref{and} and from
Figs.5 and 8 of Ref. [\onlinecite{prb2002}]:
both the stability region and the stabilization energy
for the the ground state with RS magnetic structure
are  much reduced in
contrast with the spin $S=1/2$ case considered by CNR\cite{cnr}.

\begin{figure}
\vspace{0.7cm}
 \epsfysize=60mm
 \centerline{\epsffile{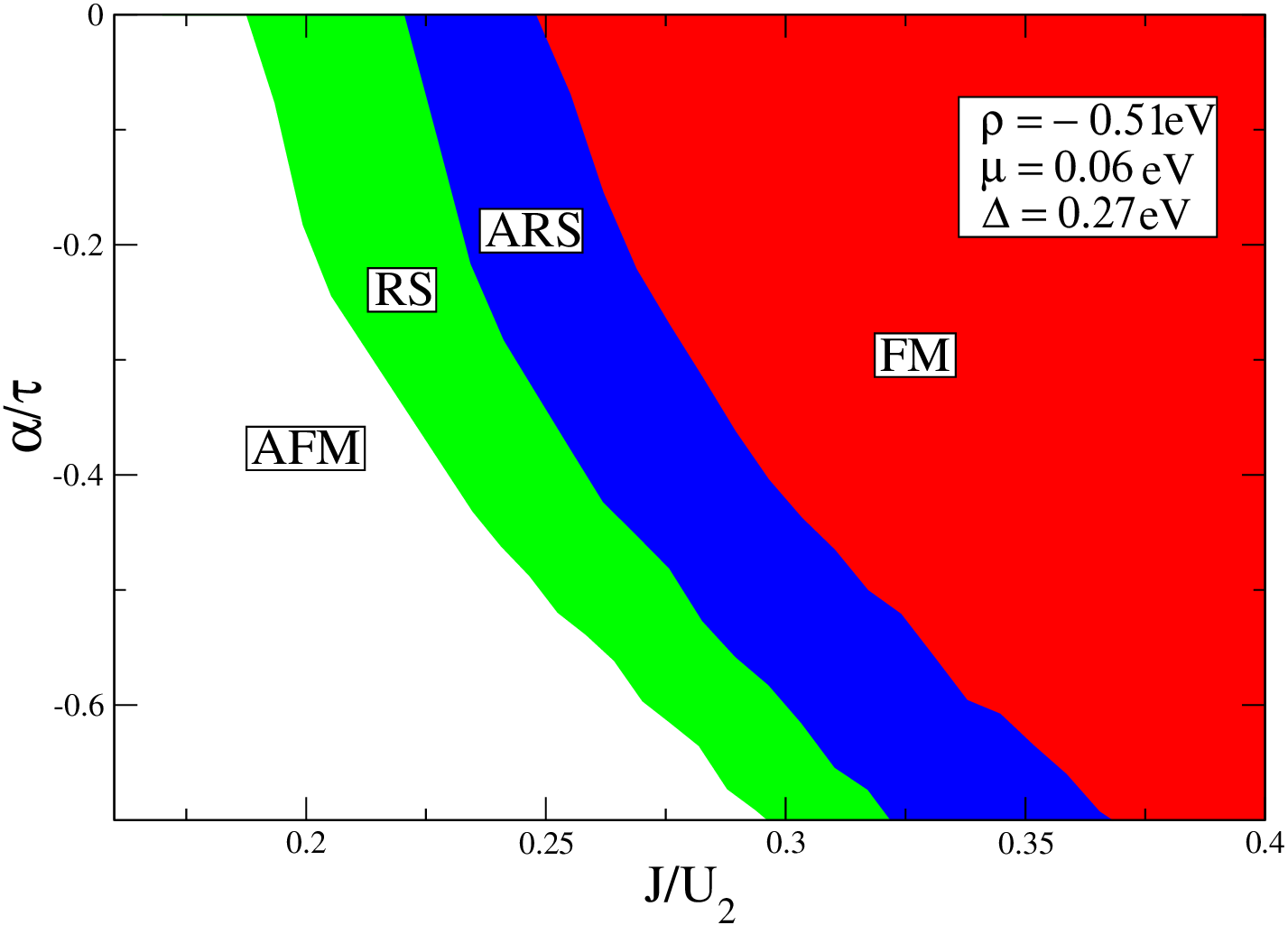}}
\caption{Phase diagram in the ($\alpha/\tau$, $J/U_2$) parameter
plane. Other hopping parameters are taken from NMTO set from Table I.
 Here AFM, FM, RS and ARS denote the corresponding type of
magnetic order. Solid lines indicate the  phase boundaries. (Color
online) } \label{phase}
\end{figure}

To illustrate more completely the properties of the effective
spin-orbital model, Eq. \ref{av}, in Fig. \ref{phase} we present its
ground-state phase diagram,  in the parameter space specified by
$\alpha/\tau$ and $J/U_2$. All other parameters are taken from the
NMTO set and we consider the trigonal splitting $\Delta_t=0.27$ eV.
>From Fig.\ref{phase} it can be seen that the ground state is
successively changed from AFM to FM magnetic phase as the Hund's
coupling increased. In order to gain maximum energy from
orbital-dependent  exchange terms, more complicated magnetic
structure are realized at the intermediate values of $J$, and
that RS structure is stabilized in a  strip between AFM and ARS
phases.

The examination  of orbital structure for various ground states,
presented on the phase diagram (Fig.\ref{phase}), has shown that
orbital configurations in  RS and ARS states are basically the same:
the orbital degeneracy is lifted and all states are occupied by only
$e_g$ electrons. For the AFM and FM phases we find a continuum of
orbital degeneracies  with any mixing angles $\theta_i$ and
$\psi_i$, depending on value of the ratio $J/U_2$ and $\alpha/\tau$.

This phase diagram is very different from  other phase diagrams,
which have been proposed for AFI of V$_2$O$_3$, i.e.  from that of
CNR model,\cite{cnr} from the molecular model of Shiina {\it et
al},\cite{mila2} from  that proposed by us in Ref. [\onlinecite{prb2002}].
The phase space occupied by RS phase is larger,  and it happens at
the realistic intermediate values of $J/U_2$.

\section{Discussion}

In this paper we have studied the ground state of the  effective
spin-orbital Hamiltonian (Eq. \ref{simpform}) with the NMTO set of
TB parameters.  We have shown that the formation of stable molecules
throughout the crystal is not favorable in this case.  For all
values of the Hund's coupling the minimization procedure with an
atomic wave function gives a lower energy than in the minimization
with a molecular wave function. The analysis of the orbital
structure of the ground state shows that the orbital degeneracy is
lifted, and that two $t_{2g}$ electrons occupy the degenerate $e_g$
doublet. We  also computed the mean-field phase diagram and showed
that the RS structure is realized at realistic intermediate values
of the Hund's coupling. The phase space of the RS phase is
significantly enlarged compared to our previous study,\cite{prb2002}
although the gain of energy compared to competing phases is again small,
of the order of $\sim 2$ meV. However, in every other aspect, the
variational orbital ground state with two $t_{2g}$ electrons occupying the
degenerate $e_g$ doublet is in disagreement with
the experimental data.

First, it contradicts the observation  of unquenched
angular momentum ($<L> \sim -0.5 \mu_B$) found by Paolasini
{\it et al}\cite{paolasini} through the use of non-resonant magnetic
x-ray scattering.  Since $<L> =0$ over a real ground state, we need
to introduce the spin-orbit coupling in the configuration $|e_g^1 e_g^2>$,
obtaining the new ground state $|e_g^+ e_g^->$,\cite{tanaka}
where $e_g^{\pm} = \frac{1}{\sqrt{2}}(\mp e_g^1 -i
e_g^2)$. Unfortunately this state still gives a zero expectation value
for the orbital moment, due to its invariance under complex
conjugation (modulo an overall sign).

Moreover it is now well established \cite{tanaka,lovesey,jdmn} that the
resonant $(1,1,1)_m$ monoclinic reflection observed by Paolasini
{\it et al}\cite{paolasini} is of magnetic origin. Specifically, it
is due to a mixture of contributions of magnetic octupole and magnetic
quadrupole moments of the V atom. The octupole moment results from a
quadrupole-quadrupole $(E_2 E_2)$ x-ray transition of even parity,
 corresponding to the operator
$[[{\bf L}\otimes \hat{\bf r}] \otimes \hat{\bf r}]^3_q$,  whereas the
quadrupole moment results from a
dipole-quadrupole $(E_1 E_2)$ x-ray transition of odd parity,
related to the spherical tensor $[{\bf L}\otimes \hat{\bf r}]^2_q$.
However, the resonant scattering amplitude (which is related to
the average values of both these operators in the $|e_g^+ e_g^-\rangle$
ground state) vanishes, again due to the invariance of
 this state under complex conjugation.

Yet another discrepancy  comes from the fact that the orbital
occupation of the $|e_g^+ e_g^-\rangle$ state disagrees with the
population analysis of the $t_{2g}$ orbitals derived by Park {\it et
al}.\cite{park00} This latter was based on polarized x-ray absorption
measurements at the $L_{2,3}$ edge of Vanadium in V$_2$O$_3$ and
provided a value of  $17\%$ for the $a_{1g}$ orbital occupancy in the
AFI phase.

Finally,  a large value of the trigonal splitting $\Delta_t$ together
with the same structure of ground state ($|e_g^1 e_g^2>$),
as found in the paramagnetic phase\cite{poteryaev} by the same NMTO
implementation of the LDA+DMFT approach,  would lead to a highly
anisotropic NMR line shape, with a  sizable quadrupole splitting in the
$^{51}$V nucleus, and to a highly anisotropic Van Vleck susceptibility.

In contrast, Rubistein\cite{rubi} found, back in 1970,  that in
the paramagnetic phase of V$_2$O$_3$  there is no detectable
anisotropy of the Knight shift and no observable quadrupole
splitting of the NMR spectrum. Even earlier, Jones\cite{jones}
performed susceptibility measurements for various crystalline orientations
with respect to the magnetic field direction and found very little  anisotropy.

We emphasize instead that all the experimental observations mentioned  above
are well explained in the framework of the molecular ground state
picture\cite{mila,prb2002}, which works especially well after the
inclusion of the spin-orbit interaction and the coupling with the
lattice degrees of freedom \cite{tanaka}.

We therefore strongly doubt the applicability of the
NMTO method to describe the physics of V$_2$O$_3$, even though the
method in itself is quite sophisticated  for band structure
calculations and is certainly in keeping with the localized,
Hubbard-type of approach to the description of correlated systems.

In fact we note that the difficulties found in the
framework of the NMTO-method when applied to V$_2$O$_3$ reflect
a general problem of all  LDA+U and LDA+DMFT calculations based on
local atomic correlations inside a single Vanadium atom.  All these
approaches lead essentially to the same kind of ground state, with two
$e_g$ electrons in spin one configuration without orbital ordering
and an almost empty $a_{1g}$ band. A similar result was found by CNR in
their second paper of the series\cite{cnr}, starting with an empirical
band structure and an unrestricted Hartree-Fock calculation,
which is very similar to the modern LDA+U scheme.

In a previous
publication\cite{sergio}, one of us has already highlighted   the
main drawbacks of implementation of DFT-based ab-initio methods to
V$_2$O$_3$. This compound is a  very peculiar strongly
correlated system, in which non-local correlations play an important
role. The molecular ground state of V$_2$O$_3$ is stabilized because
of these non-local interactions  (e.g., correlated hopping
along vertical bond), the physics of which can not be captured  within  a
method based on local interactions. The NMTO-downfolding belongs to
the same class of DFT {\it ab initio} methods and, therefore, holds
all shortcomings of local correlated approaches. Stated differently,
non-local correlations  would lead to non-local one particle
effective potentials that might substantially modify the shape and
the range of the Wannier-like functions.

One step in this direction  has  been provided by a new generation
of correlated non-local quantum chemical {\it ab initio}
calculations. In this approach a combined exact diagonalization-{\it
ab initio} method (EDABI) by Spalek {\it et al}\cite{spalek} is used
as a promising method to estimate correlated hopping. This latter is
based on a definite procedure of construction of  the many-particle
trial wave function expressed in terms of an adjustable one particle
basis, followed by the solution of a self-adjusted non-local and
non-linear wave equation obeyed by the basis functions.  We believe
that the EDABI method is particularly well suited for treating
situations in which the inter-particle correlations are not weak and
lead to non-local interactions.

Still, the major outcome of the NMTO-downfolding method is the
realization that in-plane hopping parameters are important and  a
model of weakly interacting vertical correlated molecules, which
neglects in-plane hopping, is grossly oversimplified. The interactions
between molecules, originating from in-plane hopping, are therefore not
negligible and should be included into the theory.

\end{document}